\documentclass[12pt]{article}
\usepackage{cite}
\usepackage{fix-cm}
\usepackage{color}
\usepackage[usenames,dvipsnames]{xcolor}
\usepackage{heck}
\usepackage{amsmath}
\usepackage{amsfonts}
\usepackage{amssymb}
\usepackage{graphicx}
\usepackage{multicol}
\usepackage{hyperref}
\usepackage{setspace}
\usepackage[all]{xy}
\usepackage{verbatim}
\usepackage{color}
\usepackage{epsfig}

\usepackage{calligra}
\DeclareMathAlphabet{\mathcalligra}{T1}{calligra}{m}{n}
\DeclareFontShape{T1}{calligra}{m}{n}{<->s*[2.2]callig15}{}
\usepackage{amsthm}
\input epsf.tex
\def\uv{{{}_{\rm UV}}}
\def\ir{{{}_{\rm IR}}}

\newcommand{\tsub}[1]{\ensuremath{_{\textrm{#1}}}}

\newcommand{\pd}{\partial}

\newcommand{\be}{\begin{equation}}
\newcommand{\ee}{\end{equation}}
\newcommand{\ba}{\begin{align}}
\newcommand{\ea}{\end{align}}

\newcommand{\mc}[1]{\ensuremath{\mathcal{#1}}}

\newcommand{\li}{{\bigl |}}

\newcommand{\la}{{\bigl \langle}}
\newcommand{\ra}{{\bigr \rangle}}

\def\hhat{{}}
\def\nspc{\hspace{-.5pt}}
\def\spc{\hspace{.9pt}}
\def\smpc{\hspace{.5pt}}
\def\bea{\begin{eqnarray}}
\def\eea{\end{eqnarray}}
\def\is{\! & \! = \! & \! }

\begin{document}

\addtolength{\baselineskip}{.3mm} \addtolength{\parskip}{.3mm}
\def\rr{{{}_{\rm R}}}

\def\frh{{\mbox{\small $\frac i \hbar$}}}

\def\mM{{\mbox{\tiny{$M$}}}}
\def\nN{{\mbox{\tiny{$N$}}}}
\def\ttau{\tau}
\def\aA{{\mbox{\tiny$A$}}}
\def\bB{{\mbox{\tiny$B$}}}
\def\xX{\, \underline{\nspc{x}\nspc\nspc}\,}
\def\hH{\mbox{\small$H$}}

\def\iimath{\dot{\imath}}
\addtolength{\abovedisplayskip}{.2mm}
\addtolength{\belowdisplayskip}{.2mm}
\addtolength{\baselineskip}{.3mm} \addtolength{\parskip}{.3mm}
\renewcommand{\footnotesize}{\small}
\def\dd{d} 
\def\li{{\bigl |}}
\def\ra{{\bigr\rangle}}
\def\la{{\bigl\langle}}
\def\del{\partial}

\date{{November} 2013}

\title{Geometric RG Flow}

\institution{PU}{\centerline{${}^{1}$Department of Physics,
Princeton University, Princeton, NJ 08544, USA}}

\institution{PI}{\centerline{${}^{2}$Perimeter Institute for
Theoretical Physics, 31 Caroline St. N., Waterloo, ON, N2L 2Y5,
Canada}}

\institution{Uwat}{\centerline{${}^{3}$Department of Physics \&
Astronomy,  University of Waterloo, Waterloo, Ontario N2L 3G1,
Canada}}

\authors{Steven Jackson\worksat{\PU}\footnote{e-mail: {\tt
srjackso@princeton.edu}},
Razieh Pourhasan\worksat{\PI}\worksat{\Uwat}\footnote{e-mail:
{\tt
rpourhasan@perimeterinstitute.ca}} and Herman Verlinde
\worksat{\PU}\footnote{e-mail: {\tt
verlinde@princeton.edu}}}

\abstract{We define geometric RG flow equations that specify the scale dependence of the renormalized effective action $\Gamma[g]$ and the geometric entanglement entropy ${\cal S}[x]$ of a QFT, considered as functionals of  the background metric $g$ and  the shape $x$ of the entanglement surface. We show that for QFTs with AdS duals,  
the respective flow equations  are  described by Ricci flow and mean curvature flow. For holographic theories, the diffusion rate of the RG flow is much larger, by a factor $R_{AdS}^2/\ell_s^2$,  than the RG resolution length scale. To derive our results. we employ the Hamilton-Jacobi equations that dictate the dependence 
of the total bulk action and the minimal surface area on the geometric QFT boundary data. }

\maketitle

\setlength{\textwidth}{16.2cm}
\setlength{\textheight}{22cm}

\addtolength{\abovedisplayskip}{.7mm}
\addtolength{\belowdisplayskip}{.7mm}

\section{Introduction}

QFT divergences arise due to the unlimited number of UV degrees
of
freedom. Quantities that directly measure the
number of degrees of freedom, such as the QFT effective action
$\Gamma$ receive divergent contributions from the infinitude of
short distance modes that fill space time.  Similarly, the
geometric entanglement entropy ${\cal S}$ between a region of
space $A$  and its complement contains a divergent term
proportional to the area of the entanglement surface, due to
short
distance modes that straddle the boundary of $A$ \cite{Bombelli:1986rw}. These
divergences are real and physical in any continuum QFT.

To produce finite quantities, one employs a standard
renormalization procedure: one introduces a UV regulator, adds
counterterms that cancel the divergences and then removes the
cut-off. In general, this procedure requires the introduction of
a
renormalization group scale. The renormalized quantities depend
on
the RG scale $a$ via renormalization group equations. Typically, this RG evolution
governs the scale dependence of space time
independent quantities such as the coupling constants $\phi_I$
of
the QFT lagrangian.

In this letter, we will investigate the RG evolution of the
following two quantities:
\medskip

\noindent
\parbox{16.2cm}{\addtolength{\baselineskip}{1mm}$~i)$~the
renormalized effective action $\Gamma{\!}_\rr[\phi, g\spc ]$,
defined as minus the logarithm of the CFT ${}$~~~~\,partition
function,
as a function of spatially varying couplings $\phi_I$ and
metric~$g_{\mu\nu}$,\\[1mm]
$\; ii)$ the renormalized entanglement entropy ${\cal S}_\rr[x,
a\spc ]$ of a bounded region of space,
as a {$~~~~~\,$}function of the location $x(s)$ of the
entanglement surface and RG scale $a$.}

\medskip

\noindent We will study these quantities for $d+1$-dimensional
quantum field theories that admit a holographic dual description in terms of a
weakly coupled gravitational theory in $d+2$
dimensions. Using the usual AdS/CFT dictionary and 
the Ryu-Takayanagi formula for the geometric entanglement entropy
\cite{Ryu:2006bv, Nishioka:2009un}, we
will derive that these two quantities satisfy the holographic RG flow equations
\cite{deBoer:1999xf} of the form \bea \label{flows}
\Bigl(a \frac{\delta\ }{\delta a} + \beta_n \frac{\delta\
}{\delta x_n\!}\,\Bigr)\spc{\cal S}_\rr[\spc x\spc , a\spc ] 
\, = \,  0, \!\!\!\!\!
\\[3.5mm]
\Bigl(a \frac{\delta}{\delta a} 
+ \beta_I \frac{\delta\ }{\delta \varphi_I\!} \, +
\beta_{\mu\nu}\spc \frac{\delta\ }{\! \delta g_{\mu\nu}\!} \,
\Bigr)\spc \Gamma{\!}_\rr[\spc\phi, g\spc ]\is  0.\quad\
\label{flowg}
\eea
In the second equation, we identified the generators of the RG
and Weyl rescalings via
\bea
\label{adef}
a \frac{\delta}{\delta a}
\, =\, 2 g_{\mu\nu}\nspc \frac{\delta\ }{\! \delta g_{\mu\nu}\!} .
\eea 
The above RG equations take the conventional Callan-Symanzik
form, except that we allow the RG scale to vary along the
spatial directions. Moreover,
we have included non-zero beta-functions $\beta_{\mu\nu}$ and
$\beta_n$ for the space-time metric $g_{\mu\nu}$
and for the location $x_n$ of the entanglement surface! The
respective flow equations
\bea
\label{flowb}
a\, \frac {\! \partial x_n\! }{\partial a} \, = \, \spc \beta_n,\qquad \qquad
a\, \frac {\! \partial g_{\mu\nu}\! }{\partial a}
 \is \spc 2 g_{\mu\nu} + \beta_{\mu\nu}, \qquad  \qquad
a\, \frac {\! \partial \phi_I\! }{\partial a}  
\, =\, \spc\beta_I. \eea anticipate the possibility that, in
addition to the couplings $\phi_I$, the entanglement surface and
metric also acquire a non-trivial scale dependence. We call this geometric RG
flow.

\begin{figure}[t]
\begin{center}
\includegraphics[width=.4\textwidth, trim=200 200 0
100]{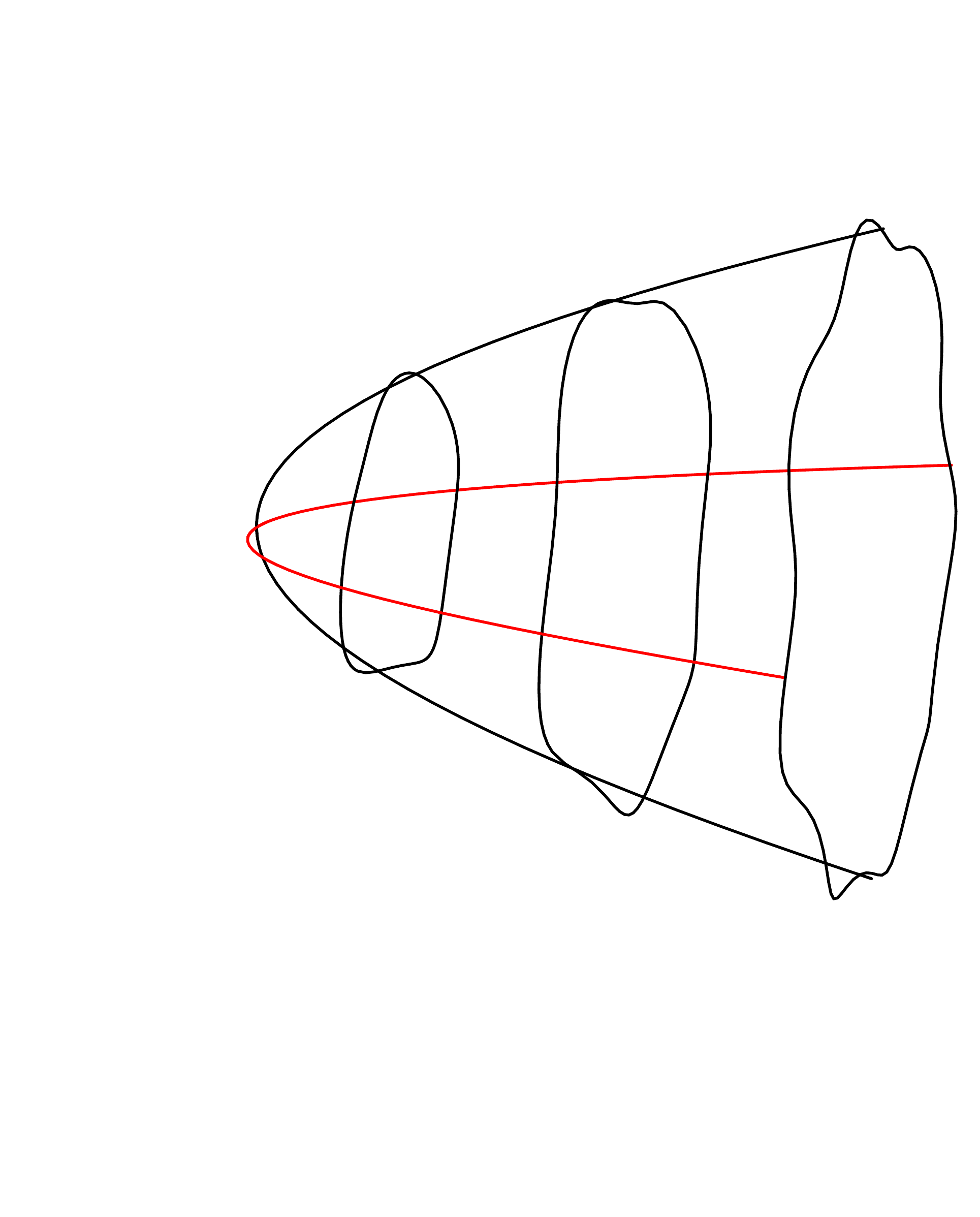}
\end{center}
\caption{The geometric RG
flow smooths out the shape of an entanglement boundary. The
smoothing effect is quantified by mean curvature flow equations.
A similar smoothing effect takes place for the background
metric, in which case the geometric RG leads to Ricci flow.
}\label{smoothingeffect}
\end{figure}

Depending on the reader's preconceptions about the meaning of RG
equations, this notion of geometric RG
flow may look either unconventional and radical, or natural and
self-evident. Normally one tends to think of the background
metric and some given entanglement surface as fixed geometric
quantities. Implicitly, one is then defining both quantities in
relation to the UV fixed point QFT. Indeed, suppose that the
metric and entanglement surface both have a fractal like shape,
with local structure divided over all possible length scales. It
is then
clear that the RG flow must have the effect of
smoothing out all UV features shorter than the RG scale, since
the
IR theory no longer notices them. This smoothing effect, depicted in fig.\ref{smoothingeffect}, is a
first indication that there should exist a notion of
geometric RG flow. 

The question remains whether the geometric RG flow can be
described via (\ref{flowb}), in terms of canonically normalized
geometric beta-functions $\beta_{\mu\nu}$ and~$\beta_n$. Naively, one would expect that
the
features subject to the flow  have size
proportional to the RG scale, so that the
associated
`beta-functions' are highly sensitive to the type of cut-off or
regulator used in setting up the RG. In the following, we will
show that this ambiguity can be avoided with the help
of
three practical restrictions. First, we will work in the regime
where geometric features such as the local curvature are small,
but non-negligible, compared to the RG cut-off scale. Moreover, we
assume that the regulator can be arranged to preserve general
covariance.
Finally, we will restrict our attention to strongly coupled
quantum field theories that admit a weakly coupled holographic
dual description.

Geometric RG flow is a natural notion within the context of
AdS/CFT. Via holography, the operation of dividing up the CFT
into
an IR and UV sector, separated by a scale $a$, amounts to
cutting
the dual AdS space-time along a spatial slice specified by some
radial location $r = r(a)$ 
\cite{Verlinde:1999xm, Heemskerk:2010hk} .  Since the bulk theory contains
gravity, the holographic RG equations that describe the radial
evolution of this slice are  defined in a diffeomorphism
invariant way. We will find that, in the weakly curved regime,
the
geometric RG flow of the metric and entanglement surface take
the
following form\footnote{In this paper, we mostly focus on RG
flow of the geometry. The spatially varying couplings $\phi_I$,
however, also take active part in the geometric RG. Their
$\beta$-functions take a similar
form of diffusion equations $\beta_I = \nu\, \nabla^2 \phi_I +
\bar{\beta}_I$ with $\nu$ and $\bar\beta_I$ scalar functions of
the scalar fields $\phi_I$.}
\bea \label{curvflow} \beta_n = \lambda \spc K_n, \qquad \ &
\qquad \beta_{\mu\nu} \, =
\, \kappa \spc\bigl( R_{\mu\nu} - \frac 1 2 R g_{\mu\nu}\bigr),&
\eea where
$R_{\mu\nu}$ and $R$ denote the Ricci tensor and scalar of the
boundary metric $g_{\mu\nu}$, and $K_n$ denotes the mean
curvature of
the entanglement surface $x_n$. Here $\lambda$ and $\kappa$ 
are scalar functions of the holographic couplings $\phi_I$,
times appropriate powers of the RG length scale $a$. For
constant $\kappa$ and $\lambda$, the geometric RG equations
(\ref{flowb})-(\ref{curvflow}) take a very familar form: the
metric evolves via {\it Ricci flow}, and the entanglement
surface
evolves via {\it mean curvature flow}.

Geometric  Ricci and mean curvature flow equations
have been actively studied in mathematics, as an analytic tool
for
proving the Poincar\'e conjecture \cite{Hamilton:1982} and related geometric
problems\cite{Colding:2011}, and in physics, in the form
of
RG equations for the coupling constants of
2-dimensional sigma models in a curved target space
\cite{Friedan:1980jf}, or with Dirichlet boundary conditions on
a
curved D-brane \cite{Leigh1989, Bakas:2007tm}.\footnote{
The connection between holographic RG and Ricci flow has
appeared
in earlier work. The geometric beta-function $\beta_{\mu\nu}$
was
first identified in \cite{Verlinde:1999xm}. In
\cite{Anderson:2011cz}, the Ricci flow was studied in the
supergravity duals of 6-D CFTs compactified on a Riemann surfacetimes ${\mathbb
R}^4$.} In both the mathematics and
physics context, it proves to be helpful to formulate the
geometric flow as a gradient flow of, respectively, the
Einstein-Hilbert action and the area or volume
functional\footnote{Here and in the following, to avoid
cluttered formulas, we will often implicitly
absorb a factor of $1/\sqrt{g}$ and $1/\sqrt{\det h}$ in the
definition of the respective variational derivatives.} \bea
R_{\mu\nu} -
\frac 1 2 R g_{\mu\nu} \, =\,  \frac{\delta\ }{\! \delta
g^{\mu\nu} }\int\!\! \sqrt{g}
\, R, \quad & & \qquad K_n \, = \, \frac{\delta \ }{\delta x^n}
\int \! \sqrt{\det h}\, , 
\eea with $h_{ij} = g_{mn} \partial_i x^m \partial_j x^n$ the
induced metric of the entanglement surface. These equations will be helpful in the
following.

In the remainder of this letter, we will derive the above flow
equations and compute the precise form of the scalar functions
$\kappa$ and $\lambda$ for a strongly coupled QFT with a given
holographic dual. Along the way, we will exhibit a direct
relation
between the geometric beta functions and the conformal anomaly.
Finally, we present the outlines of a conceptual explanation of
the geometric RG flow equations in terms the Wilsonian
interpretation of the holographic renormalization group.

\bigskip
\bigskip

\noindent
{\bf Geometric RG as a Hallmark of CFTs with AdS duals}

\smallskip

A central mystery of AdS/CFT duality is that the bulk theory
exhibits locality down to the string scale $\ell_s$,
whereas from the point of view of the boundary theory, locality
would be expected to break down at a
much larger length scale of order the AdS-radius $R_{\rm AdS}$. The
latter estimate arises because boundary to bulk propagation
causes a pointlike boundary perturbance to spread out into the
AdS-bulk over a region of size $R_{\rm AdS}$.
For this reason, one would have expected that in the Wilsonian
RG interpretation of the radial evolution,
the running RG scale, the resolution scale after coarse
graining, is set by the AdS-radius. However, the bulk gravity
theory tells us that this is incorrect: the resolution scale in
the bulk is much smaller, of order $\ell_s$.
This discrepancy lies at the heart of the holographic duality.
It also underlies the existence of geometric RG flow.

Geometric RG flow is diffusion. For example, if we linearize the
Ricci tensor $R_{\mu\nu} = \square h_{\mu\nu}$, the Ricci flow
equations literally take the form of diffusion equations. As we
will show, the
diffusion constant $\kappa$ is governed by the central charge of
the CFT, and proportional to $R_{AdS}^2$.

This interpretation of RG as diffusion sheds interesting new
light on why the resolution scale in the bulk theory can be so
small compared to the AdS radius. Also from the CFT perspective,
there are in fact {\it two} separate RG scales in the problem:
the {\it resolution} length scale $\ell_{\rm resolution}$, i.e. the
minimal length that can be resolved in the coarse grained
theory, and the {\it diffusion} length scale $\ell_{\rm
diffusion}$, that sets the diffusion rate by which the metric
and other locally varying quantities are smoothed out under the
RG flow. Holographic theories are characterized by the property
that there is a large hierarchy between the two scales
\bea
R_{\rm AdS} \sim L_{\rm diffusion} & \gg & L_{\rm resolution} \sim
\ell_{\rm string}.
\eea
In other words, the mystery of why the bulk physics is local down to
such small length scales is reversed. Our proposal is that the string
length sets the resolution scale in the Wilson RG
interpretation of the radial evolution.
The classical bulk equations of motion then imply that local geometric features
smooth out under the radial evolution with a
diffusion rate set by the AdS-radius. So from the QFT point of
view, the real unanswered question is:
\medskip

\noindent
${}$~~{\it Why do localized features of a CFT with an AdS-dual
diffuse so rapidly under RG flow?}~~${}$

\bigskip

\section{Hamilton-Jacobi}

In applications of AdS/CFT duality, many computations proceed
along the following lines: (i) one fixes some boundary
conditions
in the asymptotic AdS region, (ii) solves the bulk equations of
motion and obtains the classical solution, (iii) evaluates the
total classical bulk action, (iv) regulates and renormalizes the integral, by
subtracting any infinities that may occur. The
resulting renormalized quantity depends only on the specified
boundary conditions, and thus can be interpreted as a property
of
CFT data.

The two  quantities of interest are precisely of this type. The
QFT effective action $\Gamma{\!}_\rr[\phi,g]$ is identified withthe classical
bulk action of the gravity theory, with given
boundary values $\phi_I$ and $g_{\mu\nu}$. Similarly, the
Ryu-Takayanagi
prescription \cite{Ryu:2006bv, Nishioka:2009un}
identifies the entanglement entropy ${\cal S}[x,a]$ of a region $A$ with boundary
location specified by $x$, with the
area/volume of the
minimal RT surface $\Sigma$ in the AdS-bulk with the same
boundary
as the $A$ region, $\partial \Sigma =
\partial A$. Both quantities $\Gamma[\phi,g]$ and ${\cal
S}[x,a]$ contain UV/volume
divergences, and require a suitable holographic renormalization
procedure.

The Hamilton-Jacobi (HJ) formalism gives direct information
about
the dependence of the total classical action on the initial or
boundary conditions, for a brief nice review see
\cite{deBoer:2000cz}. It is therefore ideally suited for this
type
of problem. For the effective action $\Gamma$, this method was
introduced and worked out in
\cite{Verlinde:1999xm,deBoer:1999xf}.
Consider a $d+2$ dimensional negatively curved space-time with
metric
\bea ds^2_{d+2} 
\is G_{MN} dx^M dx^N = dr^2+ g_{\mu\nu} \spc dx^\mu dx^\nu.\eea
Here $g_{\mu\nu}$
may depend on the $d+1$ space-time coordinates $x^\mu$ and the radial coordinate
$r$. We introduce the bulk
gravity theory described by the $d+2$-dimensional action \bea
\label{sbulk}
S_{d+2} \is \int\!\! \sqrt{G} \, \bigl( R + \frac 1 2
(\partial_M \phi)^2 + V(\phi). \bigr)
\eea For definiteness, we will mostly restrict to the case that
$d= 3$. The 5-dimensional equations of motion can be written as
a
Hamiltonian system with hamiltonian $H = \int \! \sqrt{g} \;
\hhat{\cal H}$ with \bea \label{admham} \hhat{\cal H} \is\!
\pi^{\mu\nu}\pi_{\mu\nu}\! -{{1\over 3}} \pi^\mu_\mu
\pi^\nu_\nu\!
+ {{1\over 2}} \pi^I\pi_I \nspc
+ R +\nspc \frac 1 2 \nabla \phi^I \nabla \phi_I\nspc 
+ V(\phi) 
\eea
the local ADM Hamiltonian.
Here $\pi_{\mu\nu}$ and $\pi_I$ are the canonical momentum
variables conjugate to $g_{\mu\nu}$ and $\phi_I$. Let
$\Gamma[\phi,g]$ denote the classical action (\ref{sbulk}),
evaluated over a 5-d region $r< r(a)$, with boundary values
specified by $\phi$ and $g$. The HJ equation for $\Gamma$ is
obtained by setting the local ADM Hamiltonian equal to zero,
while
replacing the dual momenta by the corresponding variational
derivative of the total integrated action $\Gamma[\phi, g\spc
]$:
\bea \label{hamcon} \hhat{\cal H} = 0 \qquad \quad {\rm with} \ \
\
\ \pi_I = {\delta \spc \Gamma \over \delta \phi^I} ,& & \quad
\pi_{\mu\nu} =\frac{\delta\spc  \Gamma }{\delta g^{\mu\nu}\!}.
\eea This Hamilton constraint expresses the invariance of the
bulk
theory under reparametrizations of the radial coordinate
$r$.\footnote{In addition to the Hamiltonian constraint
(\ref{hamcon}), one also needs to impose the constraints
$\nabla^\mu \pi_{\mu\nu} + \pi_I \nabla_\nu \phi^I \, = \, 0\,$,which upon making
the same replacement as in (\ref{hamcon})
expresses the condition that $\Gamma[\phi,g]$ is invariant
underreparametrizations of the $d+1$ dimensional space-time
coordinates~$x^\mu$.}

\begin{figure}[b!]
\begin{center}
\bigskip
\bigskip

\includegraphics[width=.48\textwidth]{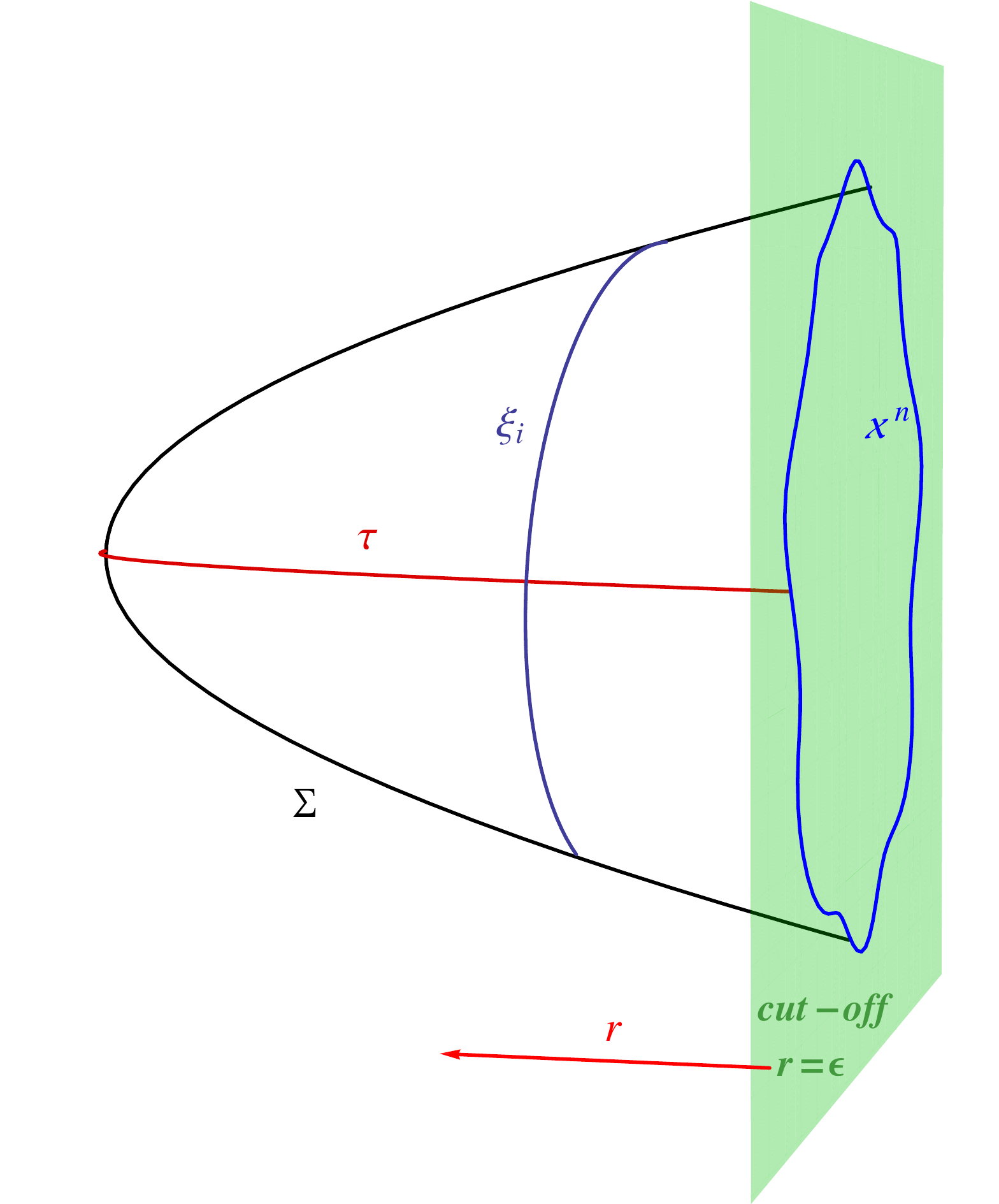}
\end{center}
\caption{The Ryu-Takayanagi minimal surface is uniquely
specified by the location $x$ of the
entanglement surface at the boundary, located at $r=
r(\epsilon)$. The total area of the RT surface, normalized by
$\frac 1 {4G_{d+2}}$, equals the holographic entanglement
entropy $\mc{S}[x,\epsilon]$ as a functional of the boundary location
$x$ and cut-off $\epsilon$, and satisfies a Hamilton-Jacobi
equation.}\label{RTsurface}
\end{figure}
Next consider a $d$-dimensional Ryu-Takayanagi minimal surface
$\Sigma$, figure (\ref{RTsurface}), suspended within a $d+1$
dimensional constant time slice with metric
\bea ds^2_{d+1} 
\is g_{\mM\nN} dx^\mM dx^\nN = dr^2+ g_{mn} \spc dx^m dx^n,\eea
where $x^\nN = (r,
x^n)$. Here {\small $N$} runs from $1$ to
$d+1$
and $n$ run from $1$ to $d$. The RT formula equates the
entanglement entropy of a CFT with a holographic dual to 1/4
times
the $d$-dimensional volume of the minimal surface $\Sigma$ \bea
\label{rtact} \mc{S} \is \frac 1 4 \int_\Sigma \! \sqrt{\det
H},\qquad \qquad H_{\aA\bB} = g_{\mM\nN} \pd_\aA x^\mM \pd_\bB
x^\nN,\eea where $\Sigma$ satisfies the asymptotic condition
$\partial \Sigma = \partial A$ at the AdS boundary. Here
$H_{\aA\bB}$ denotes the induced metric on the RT minimal
surface,
parametrized by world-volume coordinates $\xi_\aA$, with
$\mbox{\small $A$} =1, .., d$. Note that the formula
(\ref{rtact})
defines a reparametrization invariant functional of the
$\xi_\aA$
coordinates.

Again, we would like to write the Hamilton-Jacobi equation, that 
specifies the
dependence of the minimal area ${\cal S}[x]$ on
the
location of the entanglement boundary $\partial \Sigma =
\partial
A$. To do this, we first split the $d$ world-volume coordinates
$\xi_\aA = (\tau,\xi_i)$ into a radial `time coordinate' $\tau$
and $d-1$ `spatial coordinates' $\xi_i$. We can then define the
induced metric on the constant $\tau$ slice via \be \ \ \
{h}_{ij}
= g_{\mM\nN} \pd_i x^\mM \pd_j x^\nN, \ee and re-write the
volume
functional (\ref{rtact})  in first order form, as follows \bea
\label{foact} \mc{S} \is \int_\Sigma \Bigl(\pi_\nN \dot{x}^\nN -
N\bigl(g^{\mM\nN}\pi_\mM \pi_\nN + \frac 1 4 \det h \bigr) - N^i\, \pi_\nN \spc
\pd_i x^\nN\Bigr).\eea Here $\pi_\nN$ are
momenta
dual to ${x}^\nN$, and $N_\mu = (N, N^i)$ are Lagrange
multipliers
that impose the Hamilton and momentum constraints, that express
the reparametrization invariance of the original world-volume
(\ref{rtact}). The equivalence between (\ref{foact}) and
(\ref{rtact}) is easily verified.\footnote{Upon eliminating the
momentum variables via their equation of motion, one finds \be
{\cal S} = \frac 1 4 \int_\Sigma\Bigl( \frac 1 N \bigl(H_{00} +
N^iN^j H_{ij} - 2 N^i H_{i0} \bigr)- N\det H_{ij}\Bigr).\ee
Solving
for the lagrange multipliers $N_i$ gives
\be
{\cal S} = \frac 1 4 \int_\Sigma\Bigl( \frac 1 N \bigl(H_{00} -
H^{ij} H_{i0} H_{j0} \bigr)- N\det H_{ij}\Bigr).\ee
Minimizing with respect to $N$ reproduces (\ref{rtact}).} The HJ equation of the
minimal area functional ${\cal S}[x]$ is
immediately read off from the first order action (\ref{foact}), by replacing the
momentum variables in the Hamiltonian
constraint
with the corresponding variational derivatives of ${\cal S}[x]$
\be \label{pibd} \pi_r = {\partial {\cal S} \over \partial r} ,\qquad \qquad\;
\pi_{n} =\frac{\partial {\cal S} }{\partial
x^{n}}. \ee

We thus obtain the following HJ equations for ${\cal S}[x,r]$
and $\Gamma[\phi,g]$
\bea
\label{hjs} \Bigl( \frac{\partial {\cal S}}{\partial
r}\Bigr)^{\nspc
2} +\, g^{mn}\spc \frac{\partial {\cal S}\; }{\partial x^m}
\frac{\partial {\cal S} }{\partial x^n} \,
\is \frac 1 4 \spc \det h, \\[7mm]
\frac 1 {12} \Bigl(a \frac{\delta\spc \Gamma}{\delta a}\spc
\Bigr)^{\! 2} \! - 
\Bigl( \frac{\delta\, \Gamma}{\delta g^{\mu\nu}\!\!}
\,\Bigr)^{\nspc 2} - \frac 1 2 \Bigl( \frac{\delta\spc
\Gamma}{\delta \phi^I\!\!}\,\Bigr)^{\nspc 2}\!
\is  
 V(\phi)\nspc + R\nspc +\nspc \frac 1 2 (\nabla \phi)^2. 
 \label{hjg}
\eea In the second equation we used the definition (\ref{adef}).
These equations
should be read as functional identities, and can in principle be solved by means
of a derivative expansion. The
solutions are not unique, but depend on integration constants,
which can be thought of as initial or final conditions. A given
solution  uniquely specifies the radial evolution \bea
\label{flow} \textstyle \dot{g}_{\mu\nu} = 2\pi_{\mu\nu} - \frac
2
3 \pi^\lambda_\lambda \spc g_{\mu\nu}, \qquad \quad \dot{\phi}_I
=
\pi_I, \qquad \quad \dot{r} = \pi_r \qquad \quad \dot{x}_n =
\pi_n\, , \eea via the replacement of the momenta by their
`classical expectation values' (\ref{hamcon}) and (\ref{pibd}).

Equations (\ref{hjs})-(\ref{hjg}) apply equally well to the IR
part ${\cal S}_\ir$ and $\Gamma_\ir$ of both functionals, given
by
the integral over the inside AdS region $r<r(a)$, as to the UV
parts, ${\cal S}_\uv$ and $\Gamma_\uv$, given by the integral
over
outside AdS region $r>r(a)$. A UV and IR solution combined
specify a unique classical bulk trajectory.  This fact was 
explored in \cite{Verlinde:1999xm}, and forms the basis for the
interpretation of the HJ equations as RG flow of the Wilsonian
effective action. 

\def\rR{{\mbox{\footnotesize$R$}}}
\section{Hamilton-Jacobi as an RG Equation}

As a warm-up exercise for the next sections, let us write the
Hamilton-Jacobi equation for the holographic entanglement
entropy in a suggestive form, with a more direct
interpretation as an RG equation.
The idea is to split off the leading order divergent term from
the minimal area functional ${\cal S}[x,a]$.
This procedure can be seen as the first step of a
systematic derivative expansion.
For this section, we use the pure $AdS_d$ spatial metric 
\bea ds^2 \is \rR^2 \; \frac{da^2}{a^2} \spc + \spc \frac{dx^n
dx_n}{a^2}, 
\eea
with $R$ the AdS radius. With this metric, the HJ equation (\ref{hjs}) takes the form
\bea\label{eq:HJ} 
\left(a\frac{\del \mc S}{\del a}\right)^2 + \rR^2
\left(a\frac{\del \mc S}{\del x^i}\right)^2 = \frac{\rR^2}{4
a^{2d-2}} \det \hhat{h}. \eea
Here $\hhat{h}_{ij} $ has been scaled to
bring out all dependence on the holographic RG scale~$a$.

We now start with our derivative expansion: we write ${\cal S} =
{\cal S}_{\rm loc} + {\cal S}_{\rm ren}$ where the first term ${\cal
S}_{\rm loc}$ is designed to cancel the r.h.s. of the HJ
equation \eqref{eq:HJ}. We thus have
\bea\label{exp} {\cal S} = {\cal S}_{\rm loc} + {\cal S}_{\rm
ren} = \frac{\rR}{d-1} \int \frac{1}{2 a^{d-1}} \sqrt{\det
\hhat{h}} \; + \; {\cal S}_{\rm ren}.\eea
Since $\det \hhat{h}$ is independent of $a$, we immediately
verify that $\bigl(a${\large $\frac{\del {\cal S}_{\rm
loc}}{\del a}$} {$\bigr)^2 =$} {$\large \frac{R^2}{4
a^{2d-2}\!\!} $\;}{$ \det \hhat{h}$}. We see that the leading
order term ${\cal S}_{\rm loc}$ gives the expected `area contribution'.

When we vary ${\cal S}\tsub{loc}[x,a]$ with respect to the location
$x$, we obtain
\bea \frac{\delta {\cal S}\tsub{loc}}{\delta x^n} \is \frac{\rR}{\! 2
(d\nspc-\! 1)\smpc a^{d-1}\!\!}\;\, \hhat{K}_n,
\eea
where $\hhat{K}_n$ is the mean curvature of the entanglement boundary
$x$, measured with the $a$-independent metric $\hhat{h}_{\mu\nu}$. 
Plugging the expansion (\ref{exp}) into the HJ equation
\eqref{eq:HJ}, we find after a straightforward
calculation\footnote{Here we introduce the notation (c.f.
footnote 3):
$\mbox{\large $\frac{\delta}{\delta a}$} =\raisebox{2pt}{$
\frac{1}{\sqrt{\det \hhat{h}}}$} \mbox{\large $\frac{\del}{\del
a}$}$ and $\mbox{\large $\frac{\delta}{\delta x_n}$}=
\raisebox{2pt}{$ \frac{1}{\sqrt{\det \hhat{h}}}$} \mbox{\large
$\frac{\del}{\del x_n}$}$.}
\bea 
\label{srg}
a \frac{\delta {\cal S}_{\rm ren}}{\delta a} + \beta_n \spc
\frac{\delta {\cal S}_{\rm ren}}{\delta x_n} \is \, b\, K^n
K_n\, + \, \frac {a^{d-1}} {2\rR} \left(\Bigl(a \frac{\delta
{\cal S}_{\rm ren} }{\delta a}\Bigr)^2 +
\rR^2\Bigl(a\frac{\delta {\cal S}_{\rm ren}}{\delta
x}\right)^2\Bigr), \\[4mm]
\nonumber \eea
with
 \bea
 \label{coefs}
\beta_n = \frac{\rR^2 a^2 K_n}{(d\nspc-\nspc 1)}, \quad & & \quad b
= \frac{\rR^3 \spc }{4(d\nspc -\! 1)^2 a^{d-3}}\, .
\eea
 
 As promised, the new HJ equation (\ref{srg})
takes a suggestive form: it looks like an exact RG equation
that prescribes the cut-off dependence of the entanglement
entropy ${\cal S}[X,a]$ of a QFT.
The first term on the r.h.s, with pre-coefficient
$b$ given in (\ref{coefs}), indeed precisely matches with the
expected mean curvature squared term. In $d=3$, it reproduces
the correct logarithmic scale dependence prescribed by the
conformal anomaly \cite{Solodukhin:2006ic}.

The last two non-linear terms on the r.h.s. of (\ref{srg}) also
have the typical form of an exact RG, and are subleading for small
$a$. So in the following, we will mostly drop these
terms.\footnote{The quadratic terms in (\ref{srg}) do have
physical significance: they allow for topology changing
transitions along the RG flow. E.g. two disconnected components
of
the entanglement surface merge together. Indeed, the HJ
equations for the minimal surface of a holographic Wilson line
take the form of loop equations \cite{Drukker:1999zq}, of which
topology changing effects are a well-known feature. }

The renormalized term ${\cal S}_{\rm ren}$ is finite for $d<3$,
but is still logarithmically divergent at $d=3$.
In the next section we will discuss the holographic
renormalization procedure, that removes this divergence.
In dimension $d$ larger than 3, the second term ${\cal S}_{\rm
ren}$ still contains divergent terms that also need to be
removed by absorbing them into ${\cal S}_{\rm loc}$. E.g. it is
straightforward to extend the above analysis to $d=5$, and use
the HJ equation to extract the form of the 6D holographic
conformal anomaly.
\bigskip
\medskip

\section{Holographic Renormalization}

Using the previous section as a guide, we
now present a  more systematic definition of the renormalized entanglement
entropy and effective action.
In the previous two sections, we have defined ${\cal S}[x,a]$
and $\Gamma[\phi, g]$ as
the bulk contribution of classical action functionals, with the
asymptotic AdS region removed at some finite radial location $r
=
r(a)$. This truncation can be thought of as the introduction of
a
covariant UV cut-off in the dual QFT. To obtain the renormalized quantities, one
needs to remove the cut-off via holographic
renormalization \cite{Skenderis:2002wp}.

The procedure works as follows. Pick a set of renormalized
couplings $\phi^I_\rr$, metric $g_{\rr}^{\mu\nu}$ and location
$x_\rr{\!\!\!}^n$ of the entanglement boundary. These
renormalized
values are defined at some radial location inside the bulk of
AdS space, specified by the
holographic RG scale $a$. Next, evolve the holographic RG flow
equations
(\ref{flowb}) towards the UV region, until the
scale factor $a$ has attained the large value $a(\epsilon) =
\epsilon^{-1}a$.
Evaluate the functionals ${\cal S}$ and $\Gamma$ at this UV
scale
$a(\epsilon)$. The corresponding renormalized quantities are
then
defined via \bea
{\cal S}_\rr[x_\rr , a\spc ] \; \is \, \lim_{\epsilon \to 0} \;
{\cal S}_{\rm\spc finite}\bigl[\smpc x(x_\rr,\epsilon) ,
\epsilon^{-1} a\smpc \bigr], \\[2.5mm]
\Gamma{\!}_\rr[\spc \phi_\rr, \spc g_\rr ] \is \lim_{\epsilon
\to
0} \, \Gamma{\!}_{\rm\spc finite}\bigl[\smpc
\phi(\phi_\rr,\epsilon), g(g_\rr,\epsilon) \smpc \bigr]\, . \eea
Here
$x(x_\rr,\epsilon)$ denotes the value of $x$ obtained by
integrating the RG flow equations (\ref{flowb}) from $a$ to
$\epsilon^{-1} a$, with initial condition $x_\rr$, etc. The
finite
piece of the functionals is obtained by subtracting the part
that
diverges in the limit $\epsilon \to 0$ \bea {\cal S} \, = \,
{\cal
S}_{\rm div}\, + \, {\cal S}_{\rm finite} ,\quad & & \quad\Gamma
\, = \,
\Gamma{\!}_{\rm div}\, +\; \Gamma{\!}_{\rm finite}\, .
\eea
For the case $d=3$ of a 3+1-dimensional QFT, the divergent part
takes the following form
\bea
\label{sdif}
{\cal S}_{\rm div} [\smpc x\spc , a \smpc] \! \is \, \int_\Sigma
f \, \sqrt{\det{h}} \; + \;
{\cal S}_{\rm an} [\smpc x\spc , a \smpc] , 
\\[2.5mm]
 \Gamma{\!}_{\rm\spc div}[\spc  \phi, g\smpc ] \spc \is\!
 \int \!\! \sqrt{g}\spc \bigl( \spc  w +   u R\spc \bigr)  + \,
 \Gamma{\!}_{\rm an} [\smpc \phi\spc ,  g\smpc ], 
 \label{gdiv}
\eea where $f$, $w$ and $u$ are functions of the scalar
couplings
$\phi_I$, and ${\cal S}_{\rm an}$ and $\Gamma{\!}_{\rm an}$ are
logarithmically divergent terms, that carry the conformal
anomaly
of the 3+1-d QFT. To leading order in the derivative expansion,
the anomaly terms satisfy the Weyl transformation property \bea
\label{ans}
a \frac{\delta\ }{\delta a}\, 
{\cal S}_{\rm an}[x,a\spc ]   \, = \,  b \spc K^nK_n,
\qquad 
\\[3.5mm]
\quad a \frac{\delta\ }{\delta a} \spc 
\Gamma{\!}_{\rm an}[\phi,g\spc ] \, = \, c_1\spc
R^{\mu\nu}R_{\mu\nu} + c_2 \spc R^2, 
\label{ang}
\eea
where $b$, $c_1$ and $c_2$ are suitable functions of the scalar
couplings $\phi_I$.

The precise form of the six scalar functions $f$, $w$, $u$, $b$, $c_1$ and $c_2$,
as well as the higher curvature corrections to
equations (\ref{ans}) and (\ref{ang}), are determined by the
requirement that:

\smallskip

\noindent
${}$~~{The divergent terms ${\cal S}_{\rm div}$
and $\Gamma{\!}_{\rm div}$ satisfy the corresponding HJ
equations (\ref{hjs}) and (\ref{hjg}).~~}
\smallskip

 \noindent 
This requirement has a clear physical motivation: ${\cal S}_{\rm
div}$ and $\Gamma_{\rm div}$ are designed to cancel the cut-off
dependence of ${\cal S}$ and $\Gamma$, and thus should satisfy
the same radial evolution equation. We will use this property in
the next subsection, to compute the six scalar functions.

\bigskip

\section{Geometric RG Flow}

The finite parts ${\cal S}_{\rm finite}$ and $\Gamma{\!}_{\rm
finite}$ are each given by the difference of two solutions to
the
HJ equations (\ref{hjs}) and (\ref{hjg}), respectively. From
this
it is straightforward to verify that, to leading non-trivial
order
in the derivative expansion, both quantities satisfy a
holographic
RG flow equation of the form \bea \label{finsflow}
\Bigl(a \frac{\delta\ }{\delta a} + \beta_n \frac{\delta\
}{\delta x_n\!}\,\Bigr){\cal S}_{\rm finite}[x, a\smpc ] \, = \,
0 \!\! , \\[3.5mm]
\Bigl(a \frac{\delta\ }{\delta a} + \beta_{\mu\nu}\frac{\delta\
}{\delta g_{\mu\nu}\!} + \beta_I \frac{\delta\ }{\delta
\phi_I\!\!}\;\Bigr)\Gamma{\!}_{\rm finite}[\phi, g\smpc ]\is 0.
\label{fingflow} \eea In the second equation, the beta functions are obtained via
\cite{deBoer:1999xf} \bea
\gamma \, = \, a \frac {\delta\Gamma{}_{\! \rm div}}{\delta a},
\qquad \qquad 
\gamma \beta_{\mu\nu} \is \frac {\delta\Gamma{}_{\! \rm
div}}{\delta g^{\mu\nu}} , \qquad \qquad 
\gamma \beta_{I} = \frac {\delta\Gamma{}_{\! \rm div}}{\delta
\phi^{I}}\, ,
\eea where, in accord with the derivative expansion, one is
allowed to drop the anomaly term. We thus confirm that the
geometric beta function $\beta_{\mu\nu}$ takes the form of the
Ricci flow  (\ref{curvflow}). The coefficient $\kappa$  in
(\ref{curvflow}), the scalar beta function $\beta_I$ and the
other
scalar functions in (\ref{gdiv}) and (\ref{ang}) are determined
by
requiring that $\Gamma{\! }_{\rm div}$ solves the HJ
equation. One finds \cite{deBoer:1999xf} \bea
\frac 1 3 w^2\! -\! \frac 1 2 (\partial_I w)^2 \! = V, \quad & &
\qquad \kappa = \frac{6 u}{w}, \qquad \qquad \
\beta_I \spc =\spc 6\, \frac{\partial_I w}{w},  \\[3mm]
\partial^{\smpc I}\! w \, \partial_I u\nspc - \nspc \frac 1 3 w
u = 1
, \quad & & \quad\ \ c_1 = \frac{ 6 u^2}{w}, \qquad \qquad c_2
\, = \frac{2}{w}\bigl(u^2\nspc -\nspc \frac 3 2 (\partial_I u)^2
\bigr)\, .\ \
\eea

For the entanglement entropy functional ${\cal S}[x,a]$ we
proceed
in a similar way. The beta function $\beta_n$ is obtained from
${\cal S}_{\rm div}$ via \bea \label{pixrel}
\pi_r =\, \frac{\!\! \partial {\cal S}_{\rm div}\!\!}{\partial
r} \, = \gamma a\, \frac {\!\partial{\cal S}_{\rm
div}\!\!}{\partial a}, \qquad & & \qquad 
\gamma \pi_r \beta_{n} = \frac {\partial{\cal S}_{ \rm
div}}{\partial
x^{n}}\, . \eea To leading order in the derivative expansion,
one
is again allowed to drop the anomaly term. This confirms that
the
RG equation for ${\cal S}[x,a]$ takes form of the the mean
curvature flow eqn (\ref{curvflow}). The coefficient $\lambda$
in
eqn (\ref{curvflow}), and the anomaly coefficient $b$ in eqn
(\ref{ans}) are obtained by solving the following equations \bea\frac{1}{\gamma}
\, =\,a \frac {d\spc r}{d a} ,\qquad \quad
a \frac{d\ } {da} \bigl(a^{d-1} f \bigr) =
\frac{a^{d-1}}{2\gamma}, \qquad & & \
\lambda = \frac{2 f}{\gamma}, \quad \qquad b =
\frac{f^2}{\gamma}.
\eea

The above formulas specify all holographic beta functions, to
leading non-trivial order in the derivative expansion. One can
further easily verify that, for the case of pure AdS, the above
result for the anomaly coefficients $c_{1,2}$ and $b$ match with the answer
obtained via the standard technique, based on the
Fefferman-Graham expansion \cite{Skenderis:2002wp, Solodukhin:2006ic} .

To complete the derivation of the flow equation of the
renormalized quantities ${\cal S}_\rr$ and $\Gamma_{\! \rr}$, we
need
to perform the change of variables from the `bare' variables $x$ and $(\phi, g)$
to the `renormalized' variables $x_\rr$, and
($\phi_\rr, g_\rr$). As explained in more detail in
\cite{deBoer:1999xf}, this step is straightforward, since the
beta
functions transform covariantly, as vector fields. In particular\bea \beta_m(x)
\is \epsilon \, \frac {\!\partial x_m\!
}{\partial
\epsilon} \, =\, \beta_n(x_\rr) \frac{\partial x_m}{\partial
x_{{}_{\! R,\mbox{\scriptsize{$n$}}}}}. \eea This equation
expresses the basic principle that underlies the renormalization group, that a
variation in the coordinate location $x_m$ due to a
small shift in the cut-off $\epsilon$ can be compensated by an
infinitesimal adjustment of the renormalized location
$x_\rr$. Note, however, that for fixed $x_\rr$, the bare
quantity
$x_m(x_\rr,\epsilon)$ appoaches its UV fixed point value, the
location of entanglement boundary $\partial \Sigma = \partial A$ of the continuum
theory. So $\beta(x_m)$ vanishes in the limit $\epsilon \to 0$. The renormalized 
beta function $\beta_n(x_\rr)$ remains finite, however, and is proportional to the extrinsic
curvature $K_n$ of the entanglement boundary in units of the RG scale. An identical
discussion holds for the geometric beta function
$\beta_{\mu\nu}$
of the metric.

\bigskip

\section{Wilsonian Perspective}

\begin{figure}[t]
\begin{center}
\includegraphics[width=.65\textwidth]{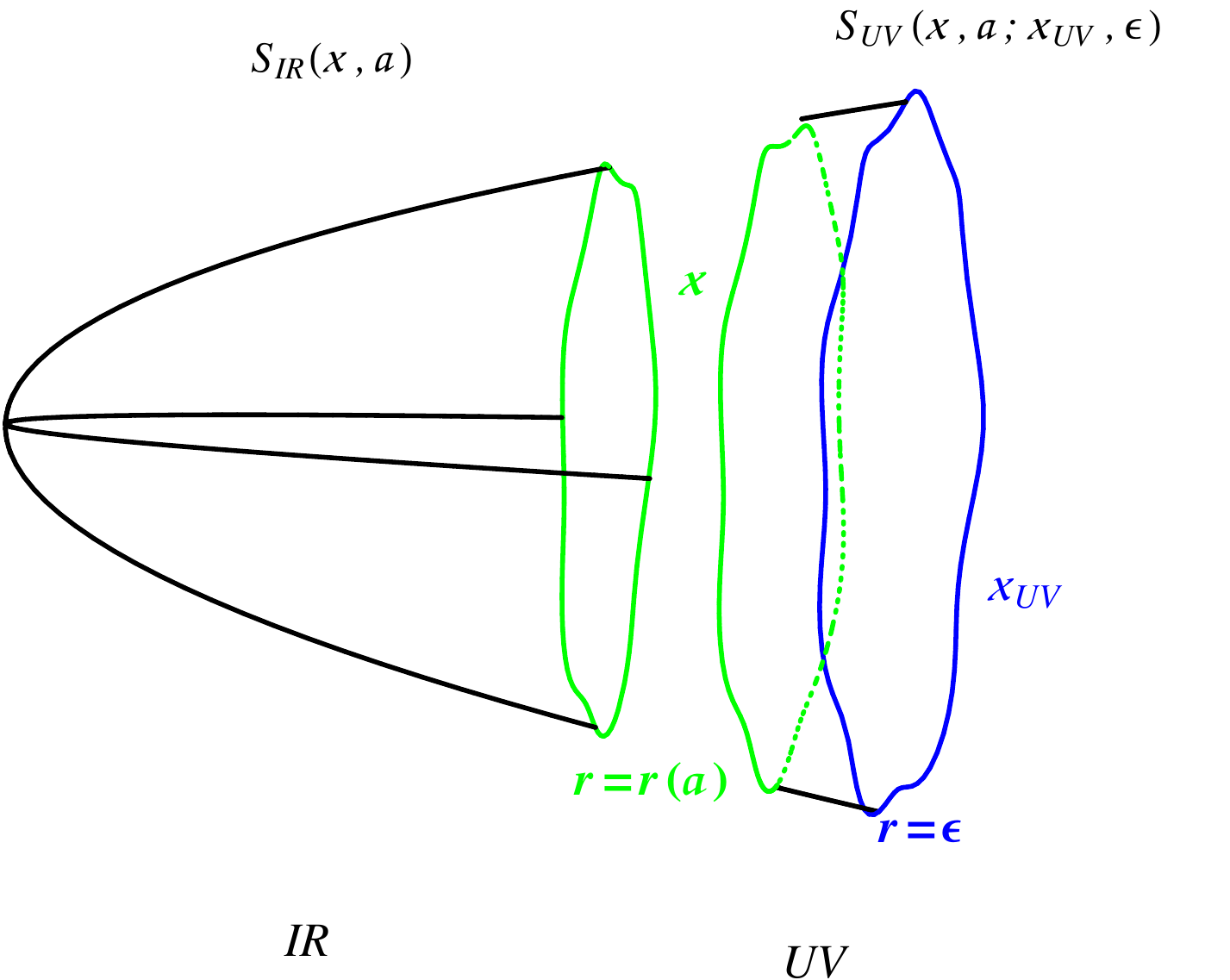}
\end{center}
\caption{The operation of dividing up the CFT into an IR and UV
sector, separated by an RG scale $a$, amounts to cutting the dual
AdS
space-time along a spatial slice specified by some radial
location
$r = r(a)$. The figure illustrates the application of this procedure to the holographic entanglement surface.}\label{AdS}
\end{figure}

To add a bit of insight into the physical origin of the
geometric
RG flow, let us recall the Wilsonian interpretation of the
holographic RG \cite{Verlinde:1999xm, Heemskerk:2010hk}. Our
discusion will be schematic.

As before, we introduce a sliding holographic RG scale by
dividing
the AdS space time into an IR region $r<r(a)$ and a UV region
$r>r(a)$. Let $S_\uv$ denote the gravity action integrated over
the UV region, and $S_\ir$ the classical action integrated over
the IR region, figure (\ref{AdS}). So both actions solve the
Hamilton-Jacobi equation (\ref{hjg}). Generalizing the standard
holographic dictionary, we can identify the bulk fields with
single trace couplings of the dual QFT \bea e^{\frh S_{{}_{\rm
IR}}[\phi,g]} 
\is \bigl\langle \,
e^{\mbox{\scriptsize $\frh\! \mbox{\,\small$\int\!$}
\bigl(\phi^I\spc  {\cal O}_I\spc + h^{\mu\nu} \spc T_{\mu\nu}
\bigr)$}\spc }\, \bigr\rangle_\ir,  \eea where the expectation
value
is presumed to be taken in the QFT with a covariant cut-off.
Here
$\hbar = 1/N$ and $h_{\mu\nu}$ denotes the metric fluctuation:
$g_{\mu\nu} = \eta_{\mu\nu} + h_{\mu\nu}$.

The partition function of the continuum QFT is obtained by
gluing
the UV and IR regions together via \cite{Verlinde:1999xm, Heemskerk:2010hk} \bea\label{zqft}
Z_{\rm QFT} \is  \int\![ d\phi \, dg] \,
\la
e^{\frh\! \mbox{\,\footnotesize$\int\!$} (\phi^I\spc {\cal
O}_I\spc + h^{\mu\nu} \spc T_{\mu\nu} )\spc }\ra_{\! \ir} \,
 e^{\frh \spc S_{{}_{\rm UV}}[\phi, g ]} \\[3mm]
 \label{zqftt}
 \is \int\![ d\phi \, dg] \, e^{\frh S_{{}_{\rm IR}}[\phi,g]\, 
+ \, \frh S_{{}_{\rm UV}}[\phi,g]} 
\eea 
In the large $N$ limit, the integral can be replaced by its
saddle point approximation. The saddle
point values of the metric and couplings are thus determined via  \cite{Verlinde:1999xm}\bea
\label{sadpt}
\frac{\delta\; }{\delta \phi^I} \bigl(S_\uv + S_\ir\bigr) \is
\frac{\delta S_\uv}{\delta \phi^I}\, + \,
\la \spc O_I \spc \ra_{\! \ir} \,  =\, 0, \\[3mm]
\frac{\delta \; }{\delta g^{\mu\nu}}\bigl(S_\uv + S_\ir\bigr)
\is \frac{\delta S_\uv}{\delta g^{\mu\nu}} \, + \,
\la \spc T_{\mu\nu} \spc \ra_{\! \ir} \, = \, 0 . \eea
From the point of view of the classical bulk gravity theory,
these are the continuity equations that ensure that the UV and
IR solutions are smoothly joined together into a single global
classical background. From the boundary QFT point of view, they
are saddle point equations that determine the value of single
trace couplings in terms of the expectation value of the dual
operators. The saddle point
value of the total classical action $S_\uv + S_\ir$ can be
identified with the quantum 1PI effective action of the QFT
\bea
\label{gammasum}
\Gamma_{\!\rr} [\phi_\rr,g_\rr] \, =\, \min_{\phi,g} \,
\bigl(\spc S_\uv[\phi,g] \spc +\spc S_\ir[\phi,g]\spc \bigr),
 \eea 
where the minimum is taken over the field values at the
holographic RG scale $r=r(a)$, while keeping the asymptotic UV boundary
conditions fixed. As explained in more  detail in
\cite{Verlinde:1999xm}, the fact that the UV and IR actions both
solve the HJ equation implies that $\Gamma_{\!\rr}$ satisfies
the geometric flow equations (\ref{flowg}) with beta-functions
determined via
\bea
\gamma(2 g_{\mu\nu} + \beta_{\mu\nu} ) \, = - \frac{\delta
S_\uv}{\delta g^{\mu\nu}} \is \la T_{\mu\nu}\ra_\ir , \\[3mm]
\gamma\beta_I \, = \, - \frac{\delta S_\uv}{\delta \phi_I } \, =
\ & & \!\!\!\!\!\!\!\!\! \!\!\!\! \la {\cal O}_I \ra_\uv .
\eea
The solution to this system of equations, up to the first few
orders in the derivative expansion, is given in the previous
section.

The Wilsonian interpretation of the holographic RG is now 
immediate. The Wilson effective action of the QFT, with the UV
cut-off in place, is defined such that after performing the
functional integral over the IR degrees of freedom, one obtains
the continuum QFT partition function $Z_{QFT}$. The appropriate Wilsonian
effective action is thus obtained simply by removing the IR
expectation values in the formula (\ref{zqft}) \bea e^{\frh
S_{\rm
eff}({\cal O}, T)}\, = \, \int\![ d\phi \, dg] \; e^{\frh\!
\mbox{\,\footnotesize$\int\!$} (\phi^I\spc  {\cal O}_I\spc +
h^{\mu\nu} \spc T_{\mu\nu} )\spc } \spc
 e^{\frh \spc S_{{}_{\rm UV}}[\phi, g ]}.
\eea 
The key point emphasized in \cite{Heemskerk:2010hk} is that
the integral over the single trace couplings on the right-hand side
implies that the effective action $S_{\rm eff}({\cal O}, T)$
contains multi-trace couplings. In particular, it contains terms that are
quadratic and higher order in the stress tensor
$T_{\mu\nu}$. 

We thus learn from the Wilsonian perspective that geometric RG flow
arises because the effective space-time metric, defined as the local
coupling constant dual to $T_{\mu\nu}$, contains operator valued
terms proportional to $T_{\mu\nu}$ itself. On a curved background, 
the stress tensor acquires a non-trivial expectation value, 
which by general covariance and to leading order is bound to be proportional to the Ricci tensor, with pre-coefficient proportional to the central charge of the CFT.
This explains why the effective background metric of the CFT evolves
via Ricci flow.

Next let us look at the geometric entanglement entropy ${\cal
S}[x,a]$ from this perspective. As we will argue, the geometric
RG flow of ${\cal
S}[x,a]$ can be derived from the geometric flow of the effective
action $\Gamma[\phi,g]$.
This is not entirely surprising, since entanglement entropy can
be represented as the limit of a QFT partition
sum on a space-time with a conical singularity \cite{calabresecardy, Nishioka:2009un}. Our discussion
will continue to be schematic.

Vacuum entanglement is a property
of the QFT vacuum state $|0\rangle$. Via the
decomposition (\ref{zqft}) of the QFT into an UV and IR sector, the state
$|\spc 0 \spc \rangle$ factorizes into an entangled sum of UV
and
IR vacuum states \bea \label{vacsum} \li \spc 0 \spc  \ra \is
\sum_{\phi,\spc g}\;  \li 0_\ir\ra_{\! {}_{\! \phi,g}} \, \li
0_\uv\ra_{\! {}_{\! \phi,g}}, \eea where $ | 0_\ir\rangle_{{}_{\!\phi,g}}$ denotes
the vacuum of the IR QFT with single trace
couplings $(\phi,g)$, while $|0_\uv\rangle_{{}_{\! \phi,g}}$
denotes the vacuum of the bulk theory in the UV region
$r>r(a)$, with IR boundary conditions $(\phi,g)$. The vacuum
state
$|0\rangle$ thus contains non-trivial UV/IR entanglement.

Rather than trying to compute the geometric entanglement entropy
of the factorized vacuum state (\ref{vacsum}),
it is more practical to look for a way to split the entanglement
entropy of the full vacuum state into an UV and IR contribution.
Let $\rho_A[x,a]$ denote the density matrix of
region $A$, bounded by the entanglement surface parametrized by
$x$, obtained by tracing over all states associated with the
geometric complement of $A$. The entanglement entropy of region
$A$ is then obtained via
\bea
{\cal S}[x,a] = - \tr\bigl( \rho_A \log \rho_A\bigr) = -
\frac{\partial\ }{\partial n} \log \tr \bigl(\rho_A^n\bigr)
\bigl|_{n=1},
\eea
where the trace is over the Hilbert space of $A$. Here we used
the standard replica trick to represent the von Neumann entropy
of $\rho_A$ as the $n\!\to \! 1$ limit of
the Renyi entropy ${\cal S}_n[x,a] = \frac 1 {1-n} \log
\tr\bigl( \rho_A^n\bigr)$. The Renyi entropy can in turn be
represented as the partition function of the QFT on a Euclidean
space-time with conical singularity, a negative deficit angle
$2\pi (1-n)$, along the entanglement surface. Since this
space-time with the conical singularity is specified by some
specific background geometry $g_{\mu\nu}$, albeit a slightly
singular one, this places us in the same setting as above. In
other words, the Renyi entropy is a quantum effective action
$\Gamma[\phi,g]$ for a particular conical background metric
$g_{\mu\nu}$ \cite{calabresecardy, Nishioka:2009un, Lewkowycz:2013nqa} .

We can thus write the Renyi entropy as a sum of a UV and an IR contribution, exactly as in eqn (\ref{gammasum}). Since the background metric data include the location of the entanglement boundary $x$, the geometric entanglement entropy involves minimization with respect to $x$
\bea
{\cal S}_\rr [x_\rr,a] \is \min_{x} \spc \bigl( {\cal S}_\ir [x,a]
+ {\cal S}_\uv[x,a]\bigr),
\eea
where $ {\cal S}_\ir$ and $ {\cal S}_\uv$ both satisfy the HJ equation.
The fact that, in accord with the Wilsonian philosophy,  this
entanglement entropy does not depend on the
RG scale $a$ is expressed in terms of an RG flow equation of the form (\ref{flows}), with exact geometric beta function $\beta_n$ given by eqns (\ref{pixrel}) with ${\cal S}_{\rm div}$ replaced by ${\cal S}_\uv$.
The Wilsonian RG evolution of the entanglement entropy and the effective action thus follow the same general pattern.

\bigskip

\bigskip

\section{Discussion}

In this paper, we derived Hamilton-Jacobi equation for the
holographic entanglement entropy formula of Ryu and Takayanagi, and proposed an interpretation in terms of a geometric RG flow in the dual QFT. We gave a unified treatment of the entanglement entropy
and the quantum effective action, and emphasized that in both cases, the RG equations include generalized beta functions that prescribe how the geometry smooths out under flow.
Here we briefly highlight some properties of the geometric RG flow and list some open ends.

\bigskip
\medskip

\noindent
{\it Universality, Diffusion and Irreversibility}

  Geometric RG  is a universal property of QFTs with weakly curved holographic duals. This class of theories have a large central charge and a gap in the 
spectrum of conformal dimensions. Combined, these two properties guarantee that there is a regime in which the RG scale dependence of the partition function and entanglement entropy can be characterized by low energy gravity and minimal surface equations. 

We have argued that geometric RG flow is a dynamical diffusion effect that is intimitely linked to but distinct from the coarse graining that drives the RG evolution. The diffusion constant is set by the AdS curvature radius $R$, which can be made arbitrarily large compared to the resolution scale, set by the string length $\ell_s$. This indicates that geometric flow is a dynamical effect. This intuition is supported by the QFT interpretation of the HJ equations in terms of the appearance of multi-trace couplings in the Wilsonian effective action.

When evolved from UV to IR,  geometric RG flow rapidly smooths out all local features  smaller
than the AdS radius. This flow is irreversible.
Suppose we can specify the geometry of space-time or of the entanglement surface at some intermediate RG scale, with local features that are smaller than the AdS radius, but still large compared to the string scale. Running the RG backwards, by integrating the bulk gravity equations of motion towards the UV,
is then like trying to run a diffusion equation backwards in time. The geometry will develop a singularity 
at some finite RG time, much before reaching the UV boundary.

\bigskip
\medskip

\noindent
{\it RG Fixed Points  and Topology}

With conventional RG, one often considers flows from a UV fixed point to some IR fixed point. 
The fixed points of the curvature flow (\ref{curvflow}) are Ricci flat space-times and minimal surfaces with zero extrinsic curvature. To get a non-trivial flow between different fixed points, one would need
to consider situations with non-trivial topology. One could  also consider modifying the equations by absorbing a non-trivial scaling factor into the RG step, so that the fixed geometries are allowed to have non-zero constant curvature. 

When the entanglement region consists of two or more disconnected but sufficiently proximate regions  in the UV, it is possible that while evolving the RG towards the IR, the different components of the entanglement boundary merge together. If in addition, one also considers space-time geometries with non-trivial topology, one can easily imagine setting up RG flows between different topologically distinct fixed points geometries.

\bigskip
\medskip

\noindent
{\it QFT Derivation?}

An outstanding challenge is to find a pure QFT derivation of the geometric RG flow equations.
This may seem impossible, because geometric flow is a property of strongly coupled quantum field theories with holographic duals. In the case of AdS${}_3$/CFT${}_2$, however, it is known how to derive the 2+1-D bulk gravity equations of motion (in Hamilon-Jacobi form) from the 2-D conformal anomaly and Virasoro Ward-identities of the boundary CFT. In that case, it is possible to give a CFT proof of the geometric RG flow equations. However, due to the low dimension  and absence of intrinsic geometry of the entanglement boundaries, the Ricci and mean curvature flow degenerate into a rather trivial form. It may still be worth trying,  via clever use of various techniques (such as Ward identities of the stress tensor, conformal anomaly, cone geometries, conformal perturbation theory, etc) to find a derivation of the geometric flow from a CFT in $d\geq 2$. A somewhat tantalizing hint is that the beta functions of the geometric RG look identical to those of an (open) string world sheet sigma model. This suggests that an open string represention of the QFT may be a helpful tool \cite{Polyakov:1993tp}.

\bigskip

\section*{Acknowledgements}

We would like to thank Jan de Boer, Aitor Lewkowycz, Shiraz Minwalla, Rob Myers, Tatsuma Nishioka, Guilherme Pimentel, Sasha Polyakov, Misha Smolkin, Joe Polchinski and Erik Verlinde
 for useful discussions and comments. The research of S.J. and H.V. is supported by NSF grant PHY-1314198. R.P. is supported by NSERC of Canada
and
in part by Perimeter Institute for Theoretical Physics. Research at Perimeter
Institute is supported by the Government of Canada
through Industry Canada and by the Province of Ontario through
the
Ministry of Research and Innovation. R.P. would also like to
thank
Frankfurt Institute for Advanced Studies whilst this work was
initiated and Princeton University where it was finalized for
their
hospitality and financial support.

\addtolength{\baselineskip}{-.3mm}


\begin{thebibliography}{99}
\bibitem{Bombelli:1986rw} 
  L.~Bombelli, R.~K.~Koul, J.~Lee and R.~D.~Sorkin,
  Phys.\ Rev.\ D {\bf 34}, 373 (1986)'
  M.~Srednicki,
  Phys.\ Rev.\ Lett.\  {\bf 71}, 666 (1993)
  [hep-th/9303048].
  
\bibitem{Ryu:2006bv}
S.~Ryu and T.~Takayanagi, {\it {Holographic derivation of entanglement entropy
  from AdS/CFT}},  {\em Phys.Rev.Lett.} {\bf 96} (2006) 181602.

  \bibitem{Nishioka:2009un} 
  T.~Nishioka, S.~Ryu and T.~Takayanagi,
  J.\ Phys.\ A {\bf 42}, 504008 (2009)
  [arXiv:0905.0932 [hep-th]].
  
\bibitem{deBoer:1999xf}
J.~de~Boer, E.~P. Verlinde, and H.~L. Verlinde, {\it {On the holographic
  renormalization group}},  {\em JHEP} {\bf 0008} (2000) 003
  
\bibitem{deBoer:2000cz}  J.~de~Boer, {\it {The Holographic renormalization group}},  {\em Fortsch.Phys.}
  {\bf 49} (2001) 339--358, [\href{http://xxx.lanl.gov/abs/hep-th/0101026}{{\tt
  hep-th/0101026}}].

  \bibitem{Skenderis:2002wp} 
  K.~Skenderis,
  Class.\ Quant.\ Grav.\  {\bf 19}, 5849 (2002)
  [hep-th/0209067];
  I.~Papadimitriou and K.~Skenderis,
  hep-th/0404176.
 
\bibitem{Verlinde:1999xm}
E.~P. Verlinde and H.~L. Verlinde, {\it {RG flow, gravity and the cosmological
  constant}},  {\em JHEP} {\bf 0005} (2000) 034,
  [\href{http://xxx.lanl.gov/abs/hep-th/9912018}{{\tt hep-th/9912018}}].

\bibitem{Heemskerk:2010hk} 
  I.~Heemskerk and J.~Polchinski,
  JHEP {\bf 1106}, 031 (2011);
  T.~Faulkner, H.~Liu and M.~Rangamani,
  JHEP {\bf 1108}, 051 (2011); D.~Radicevic,
  JHEP {\bf 1112}, 023 (2011)
  [arXiv:1105.5825 [hep-th]]; S.~-S.~Lee,
  arXiv:1305.3908 [hep-th].
 
\bibitem{Hamilton:1982}
R.~S. Hamilton, {\it {Three-manifolds with positive Ricci curvature}},  {\em
  J.Diff.Geom.} {\bf 17} (1982) 255--306;
G.~Perelman, {\it {The Entropy formula for the Ricci flow and its geometric
  applications}},  \href{http://xxx.lanl.gov/abs/math/0211159}{{\tt
  math/0211159}};
{\it {Ricci flow with surgery on three-manifolds}},
  \href{http://xxx.lanl.gov/abs/math/0303109}{{\tt math/0303109}}.

\bibitem{Colding:2011}
T.~H. Colding and W.~P. Minicozzi~II,  {\em Ann. of Math.} {\bf 175} (2012) 755--833,

\bibitem{Friedan:1980jf}
D.~Friedan,  {\em
  Phys.Rev.Lett.} {\bf 45} (1980) 1057.
\bibitem{Leigh1989} 
  R.~G.~Leigh,
  Mod.\ Phys.\ Lett.\ A {\bf 4}, 2767 (1989).
  
  
  
\bibitem{Bakas:2007tm}
I.~Bakas and C.~Sourdis, {\em JHEP} {\bf 0706} (2007) 057.

  
\bibitem{Anderson:2011cz}
M.~T. Anderson, C.~Beem, N.~Bobev, and L.~Rastelli,  {\em Commun.Math.Phys.} {\bf 318} (2013) 429--471,
  [\href{http://xxx.lanl.gov/abs/1109.3724}{{\tt arXiv:1109.3724}}].

  \bibitem{Solodukhin:2006ic} 
  S.~N.~Solodukhin,
  Phys.\ Lett.\ B {\bf 646}, 268 (2007)
  [hep-th/0609045];
  Phys.\ Lett.\ B {\bf 665}, 305 (2008)
  [arXiv:0802.3117];  D.~V.~Fursaev,
  JHEP {\bf 0609}, 018 (2006);
  D.~V.~Fursaev, A.~Patrushev and S.~N.~Solodukhin,
  arXiv:1306.4000.
\bibitem{Drukker:1999zq} 
  N.~Drukker, D.~J.~Gross and H.~Ooguri,
  Phys.\ Rev.\ D {\bf 60}, 125006 (1999)
  [hep-th/9904191];
  A.~M.~Polyakov and V.~S.~Rychkov,
  Nucl.\ Phys.\ B {\bf 581}, 116 (2000)
  [hep-th/0002106];
  Nucl.\ Phys.\ B {\bf 594}, 272 (2001)
  [hep-th/0005173].
  \bibitem{calabresecardy} P. Calabrese and J. L. Cardy, “Entanglement entropy and quantum field theory: A non- technical introduction,” Int. J. Quant. Inf. 4 (2006) 429 [arXiv:quant-ph/0505193].
  \bibitem{Lewkowycz:2013nqa} 
  A.~Lewkowycz and J.~Maldacena,
  ``Generalized gravitational entropy,''
  JHEP {\bf 1308}, 090 (2013)
  [arXiv:1304.4926 [hep-th]].
  \bibitem{Verlinde:1989ua} 
  H.~L.~Verlinde,
  Nucl.\ Phys.\ B {\bf 337}, 652 (1990);
  L.~Freidel,
  arXiv:0804.0632 [hep-th].
 \bibitem{Polyakov:1993tp} 
  A.~M.~Polyakov,
  ``A Few projects in string theory,''
  hep-th/9304146.
  J.~Khoury and H.~L.~Verlinde,
  Adv.\ Theor.\ Math.\ Phys.\  {\bf 3}, 1893 (1999)
  [hep-th/0001056].
\end{thebibliography}

\end{document}